\documentclass[sigconf]{acmart}
\AtBeginDocument{%
  \providecommand\BibTeX{{%
    \normalfont B\kern-0.5em{\scshape i\kern-0.25em b}\kern-0.8em\TeX}}}

\copyrightyear{2023}
\acmYear{2023}
\setcopyright{rightsretained}
\acmConference[CIKM '23]{Proceedings of the 32nd ACM International Conference on Information and Knowledge Management}{October 21--25, 2023}{Birmingham, United Kingdom}
\acmBooktitle{Proceedings of the 32nd ACM International Conference on Information and Knowledge Management (CIKM '23), October 21--25, 2023, Birmingham, United Kingdom}
\acmDOI{10.1145/3583780.3615165}
\acmISBN{979-8-4007-0124-5/23/10}

\usepackage{booktabs}
\usepackage{enumitem}
\usepackage{mathtools}
\usepackage{array}
\usepackage{multirow}
\usepackage{xcolor,soul}
\usepackage{arydshln}

\newcommand{\bftab}{\fontseries{b}\selectfont}
\newcommand{\ittab}{\fontshape{it}\selectfont}

\newcolumntype{H}{>{\setbox0=\hbox\bgroup}c<{\egroup}@{}}

\definecolor{lightgray}{RGB}{225, 225, 225}

\DeclareRobustCommand{\hlgray}[1]{{\sethlcolor{lightgray}\hl{#1}}}

\newcolumntype{P}[1]{>{\centering\arraybackslash}p{#1}}

\begin{document}

\title[Counterfactual Graph Augmentation for Consumer Unfairness Mitigation in Recommender Systems]{Counterfactual Graph Augmentation for \\Consumer Unfairness Mitigation in Recommender Systems}


\author{Ludovico Boratto}
\orcid{0000-0002-6053-3015}
\affiliation{%
  \institution{University of Cagliari}
  \streetaddress{Via Ospedale, 72}
  \city{Cagliari}
  \country{Italy}
}
\email{ludovico.boratto@acm.org}

\author{Francesco Fabbri}
\orcid{0000-0002-9631-1799}
\affiliation{%
  \institution{Spotify}
  \streetaddress{}
  \city{Barcelona}
  \country{Spain}
}
\email{francescof@spotify.com}

\author{Gianni Fenu}
\orcid{0000-0003-4668-2476}
\affiliation{%
  \institution{University of Cagliari}
  \streetaddress{Via Ospedale, 72}
  \city{Cagliari}
  \country{Italy}
}
\email{fenu@unica.it}

\author{Mirko Marras}
\orcid{0000-0003-1989-6057}
\affiliation{%
  \institution{University of Cagliari}
  \streetaddress{Via Ospedale, 72}
  \city{Cagliari}
  \country{Italy}
}
\email{mirko.marras@acm.org}

\author{Giacomo Medda}
\orcid{0000-0002-1300-1876}
\affiliation{%
  \institution{University of Cagliari}
  \streetaddress{Via Ospedale, 72}
  \city{Cagliari}
  \country{Italy}
}
\email{giacomo.medda@unica.it}

\renewcommand{\shortauthors}{Boratto et al.}

\begin{abstract}
  In recommendation literature, explainability and fairness are becoming two prominent perspectives to consider. However, prior works have mostly addressed them separately, for instance by explaining to consumers why a certain item was recommended or mitigating disparate impacts in recommendation utility. None of them has leveraged explainability techniques to inform unfairness mitigation. In this paper, we propose an approach that relies on counterfactual explanations to augment the set of user-item interactions, such that using them while inferring recommendations leads to fairer outcomes. 
  {\color{black} Modeling user-item interactions as a bipartite graph, our approach augments the latter by identifying new user-item edges that not only can explain the original unfairness by design, but can also mitigate it.}
  {\color{black} Experiments on two public data sets show that our approach effectively leads to a better trade-off between fairness and recommendation utility compared with state-of-the-art mitigation procedures.} We further analyze the characteristics of added edges to highlight key unfairness patterns. Source code available at \url{https://github.com/jackmedda/RS-BGExplainer/tree/cikm2023}.
\end{abstract}





\maketitle

\section{Introduction}


Current research in recommender systems is increasingly focusing on beyond-utility perspectives, such as explainability~\cite{ZhangC20} and fairness~\cite{WanngMZLM22}, also in response to the recently issued regulations~\cite{capAI}.
However, such perspectives are usually considered separately. 
For instance, research into explainability has focused on merely justifying why a certain item has been included within the recommended list~\cite{DBLP:conf/pkdd/ChenSilvestriPKDD2022,DBLP:conf/pkdd/KangLB21}, often without inspecting whether and why adding that item might lead to disparate impacts on demographic groups. 
Conversely, existing methods to mitigate unfairness have often relied on mathematical formulations of fairness principles, but rarely informed from explanatory analyses on such unfairness~\cite{DBLP:conf/fat/BurkeSO18,DBLP:conf/fat/EkstrandTAEAMP18,DBLP:conf/www/LiCFGZ21,DBLP:conf/fedcsis/FrischLG21}.
Concerted efforts towards explaining unfairness in recommendation have been recently made ~\cite{DBLP:journals/ipm/DeldjooBN21,DBLP:conf/sigir/GeTZXL0FGLZ22}. Unfortunately, none of them has led to a mitigation procedure that leverages the identified explanations to mitigate the measured unfairness.
A first attempt to inform a mitigation procedure through explanation techniques was proposed by \cite{DBLP:conf/aaai/DongW0LL23}.
However, their method operates on Graph Neural Networks (GNNs) to detect the graph nodes affecting unfairness in classification tasks, limiting its adoption to networks of user-item interactions and to the recommendation task in general.

In this paper, we propose an approach that augments a user-item interactions graph to counteract consumer
unfairness~\cite{10.1007/978-3-030-99736-6_37} across demographic groups in recommendation.
Following works that used counterfactual techniques in GNNs \cite{DBLP:conf/aistats/LucicHTRS22,DBLP:conf/pkdd/KangLB21,DBLP:conf/pkdd/ChenSilvestriPKDD2022,DBLP:conf/kdd/YuanTHJ20}, our framework aims to modify the top-$k$ lists generated by a GNN-based recommender system by adding edges to the graph used in the model inference step, such that the altered recommended lists are fairer across demographic groups of consumers.
The augmentation mechanism is guided by a two-term loss function that selects the minimum set of edges to solve the targeted task.
{\color{black} Specifically, we assume that the actions of the users in a demographic group led the model to \emph{advantage} them.
Thus, we hypothesize a \emph{counterfactual} world where the \emph{disadvantaged} users can benefit from new edges to improve their recommendation utility.}
If our approach accomplishes this task, the additional edges represent an explanation (i) of the fairness-related knowledge missing from the non-augmented graph, and (ii) of the underlying mitigation process of our method.
On two public real-world data sets, our method demonstrates greater reliability than state-of-the-art (SOTA) techniques in mitigating consumer unfairness.
We also describe the original unfairness by analyzing the user and item nodes involved in the added edges.

\section{Related Work}

Despite the abundance of consumer unfairness mitigation procedures in recommendation~\cite{10.1007/978-3-030-99736-6_37,DBLP:conf/fat/EkstrandTAEAMP18,DBLP:conf/fat/BurkeSO18, DBLP:conf/fedcsis/FrischLG21,DBLP:conf/www/LiCFGZ21,DBLP:conf/sigir/LiCXGZ21,DBLP:conf/aaai/WuWWH021,10.1145/3564285},
the aspects that lead such techniques to successfully improve fairness still remain nebulous.
It is also uncertain whether unfairness explainability methods in recommendation \cite{DBLP:journals/ipm/DeldjooBN21,DBLP:conf/sigir/GeTZXL0FGLZ22} could be leveraged to mitigate and not only explain the issue.
In GNNs literature, emphasis was put on improving explainability \cite{DBLP:conf/sigir/BalloccuBFM22, DBLP:journals/corr/abs-2202-06466,DBLP:conf/wsdm/GhazimatinBRW20,DBLP:conf/www/GhazimatinPRW21} and fairness \cite{DBLP:conf/uai/AgarwalLZ21,DBLP:conf/wsdm/MaGWYZL22,DBLP:conf/www/WangLLW22} for several downstream tasks.
\cite{DBLP:conf/aistats/LucicHTRS22,DBLP:conf/pkdd/KangLB21,DBLP:conf/kdd/YuanTHJ20,DBLP:conf/pkdd/ChenSilvestriPKDD2022} adopted counterfactual methods to modify (e.g. perturbation) the graph topology, but only a few of them were studied for recommendation \cite{DBLP:conf/pkdd/ChenSilvestriPKDD2022,DBLP:conf/pkdd/KangLB21}, and not necessarily considering unfairness issues.
The information provided by explanation methods on graphs was leveraged for additional tasks in \cite{DBLP:conf/log/GiunchigliaSGA22,DBLP:conf/aaai/DongW0LL23}.
Despite fairness was contemplated in \cite{DBLP:conf/aaai/DongW0LL23},
such works were devised for classification purposes on monopartite graphs.

\section{Methodology} \label{sec:method}

\subsection{Problem Formulation}

\subsubsection{Recommendation Task} \label{subsubsec:rec_task}
The goal is to predict whether or the level of interest that a user $u \in U$ may have for an unseen item $i \in I$.
The interactions between users and items can be represented by a bipartite graph $G = (U, I, E)$, where $U \cup I$ ($n = |U| + |I|$) is the set of nodes and $E$ is the set of edges connecting such nodes.
Let $A$ be a $n \times n$ adjacency matrix representing $G$, missing links can be predicted by any GNN, defined as $f(A, W) \rightarrow \hat{R} \in \mathbb{R}^{|U| \times |I|}$, where $\hat{R}_{u,i}$ represents the linking probability between $u$ and $i$, and $f$ is parameterized by the weight matrix $W$.
{\color{black} A list of the $k$ items with the highest probability in $\hat{R}_u$ is recommended to each user $u$.}

\subsubsection{Mitigation Task} \label{subsubsec:mitig_task}

{\color{black} Our goal is to make $f$ produce altered yet fairer recommendations.}
To this end, we leverage counterfactual reasoning techniques \cite{DBLP:conf/aistats/LucicHTRS22, DBLP:conf/pkdd/KangLB21} to generate a minimally augmented version of $A$, i.e. $\tilde{A}$, such that the utility estimates across consumers' groups are not systematically different when $f$ uses $\tilde{A}$ instead of $A$ during inference.
We base our fairness notion on \emph{demographic parity}, emphasized in top-$k$ recommendation by prior work~\cite{DBLP:journals/ipm/BorattoFMM23,10.1145/3564285,DBLP:conf/www/LiCFGZ21}.
According to it, we aim to minimize the following loss function:
\begin{equation} \label{eq:loss}
    \mathcal{L}(A, \tilde{A}) = \mathcal{L}_{fair}(A, f(\tilde{A}, W)) + \mathcal{L}_{dist}(A, \tilde{A})
\end{equation}
$\mathcal{L}_{fair}$ quantifies fairness, by using the operationalized demographic parity function; $\mathcal{L}_{dist}$ controls the distance between $A$ and $\tilde{A}$.

\subsection{Graph Augmentation} \label{subsec:graph_augment}

\subsubsection{Augmentation Mechanism} \label{subsubsec:augment_mech}
Similarly to \cite{DBLP:conf/aistats/LucicHTRS22}, {\color{black} we developed a mechanism tailored for bipartite graphs to modify their topology.
\cite{DBLP:conf/aistats/LucicHTRS22} uses a matrix $P$ to perturb A, i.e. $\tilde{A} = P \odot A$
, while our method reduces the memory usage by augmenting $A$ with a predefined set of $B$ edges, based on a vector $p$.}
{\color{black} We generate $\tilde{A}$ by augmenting the missing entries in $A$ with the entries in $p$ through a function $h: \mathbb{N}^{|U|} \times \mathbb{N}^{|I|} \rightarrow \mathbb{N}$,} that maps the 2D indices ($u$, $i$) of $A$ into a 1D index for $p$, such that an edge $\tilde{A}_{u,i}$ is added if $p_{h(u,i)} = 1$.
Formally:
\begin{equation} \label{eq:augment}
    \begin{aligned}
        \quad & \tilde{A}_{u,i} =
        \begin{cases}
            {p}_{h(u,i)} & if \: h(u,i) \in \mathbb{N}_{<B} \\
            A_{u,i} & otherwise
        \end{cases} \\
    \end{aligned}
\end{equation}
$p$ is derived from a real valued vector $\hat{p}$, as done in \cite{DBLP:conf/aistats/LucicHTRS22,DBLP:conf/cvpr/SrinivasSB17}, by applying a sigmoid transformation before rounding values $\ge 0.5$ to $1$ and values $< 0.5$ to 0.
We initialize $\hat{p}_i = -5, \forall i \in [0, B)$, such that $\hat{p}_i \approx 0$ after the sigmoid transformation and it is guaranteed $\tilde{A} = A$.

\subsubsection{Augmented Graph Generation} \label{subsubsec:augment_gener}

{\color{black} The core process of the augmented graph generation is carried out by an extended version of $f$, denoted as $\tilde{f}(\tilde{A}, W; \hat{p}) \rightarrow \tilde{R}$, that shares the same implementation with $f$, but $W$ remains constant as an additional input.
Differently from $f$, $\tilde{f}$ (i) leverages the vector $\hat{p}$ as parameter to perform the augmentation mechanism, resulting in $\tilde{A}$, (ii) retrieves a matrix $\tilde{R}$ with linking probabilities altered by the usage of $\tilde{A}$ during inference, and (iii) updates $\hat{p}$ according to \eqref{eq:loss}.
In other words, $\tilde{f}$ performs an iterative process to augment the graph until a fairness requirement estimated on $\tilde{R}$ is satisfied.
The final augmented matrix $\tilde{A}$ represents a distorted version of $A$ in a counterfactual world.
If $\tilde{A}$ makes $f$ produce recommendations with fairer estimates on the perturbation set, $\tilde{A}$ provides a \emph{counterfactual explanation} of the prior unfairness on the same set, similarly to \cite{DBLP:conf/sigir/GeTZXL0FGLZ22}.
We expect that adopting $\tilde{A}$ could also mitigate the unfairness on the evaluation set.}


\subsubsection{Loss Function Optimization} \label{subsubsec:loss_optim}

{\color{black} Let $\mathcal{G}$ be the set of demographic groups, we operationalize $\mathcal{L}_{fair}$ (see \eqref{eq:loss}) as in recent works~\cite{10.1007/978-3-030-99736-6_37,10.1145/3564285}:}
\begin{equation} \label{eq:fairloss}
    \fontsize{8.9pt}{11pt}
    \mathcal{L}_{fair}(A, \tilde{R}) = \frac{1}{\binom{|\mathcal{G}|}{2}}\sum_{1 \leq i < j \leq |\mathcal{G}|}{\left\Vert S(\tilde{R}^{\mathcal{G}_i}, A^{\mathcal{G}_i}) - S(\tilde{R}^{\mathcal{G}_j}, A^{\mathcal{G}_j}) \right\Vert^2_2}
\end{equation}
where $A^{\mathcal{G}_i}$ and $\tilde{R}^{\mathcal{G}_i}$ denote the adjacency and linking probabilities sub-matrices related to the users in the $i$-th group, $S$ is a recommendation utility metric.
We selected Normalized Discounted Cumulative Gain (NDCG) as the latter in the evaluation phase, but, due to its non-differentiability, we adopted an approximated version \cite{10.1145/3564285,DBLP:journals/ir/QinLL10}\footnote{We use the TensorFlow implementation, called \texttt{ApproxNDCGLoss}.}, i.e. $\widehat{NDCG}$, to optimize $\mathcal{L}_{fair}$.
We denote the data subset from which the ground truth labels are taken to measure $\widehat{NDCG}$ during the optimization process as the \emph{perturbation} set.
With focus on a binary setting as prior studies \cite{DBLP:conf/www/LiCFGZ21,DBLP:journals/ipm/AshokanH21,DBLP:conf/fat/KamishimaAAS18}, we define the subsets $U_D = \{u \in U \: | \: u \in \mathcal{G}_D \}$ and $U_A = \{u \in U \: | \: u \in \mathcal{G}_A \}$, where $\mathcal{G}_D, \mathcal{G}_A$ are the \emph{disadvantaged} and \emph{advantaged} groups respectively.
The group with lower (higher) utility on the \emph{perturbation} set is denoted as disadvantaged (advantaged).
Our approach aims to increase the utility of the disadvantaged group (not to reduce the advantaged group's one).
Edges are only added to user nodes in $U_D$.
{\color{black} Conversely, $\mathcal{L}_{dist}$ (see \eqref{eq:loss}) can be any differentiable distance function~\cite{DBLP:conf/aistats/LucicHTRS22}:}
\begin{equation}
    \mathcal{L}_{dist} = \beta \frac{1}{2}\sigma\left(\sum_{i, j}{\left\Vert\tilde{A}_{i,j} - A_{i,j}\right\Vert^2_2}\right)
\end{equation}
where $\sigma$ is a sigmoid function\footnote{We use $\sigma(x) = |x|/(1 + |x|)$ since it grows slower than other functions as $x$ increases.} to bound $\mathcal{L}_{dist}$ in the range $[0,1]$ as $\mathcal{L}_{fair}$, $\beta$ is a scaling factor that balances the two losses.
We set $\beta = 0.5$ to give more importance to the mitigation task, i.e. $\mathcal{L}_{fair}$.

\subsection{Sampling Policies} \label{subsec:sampl_policies}

Even if the loss function in \eqref{eq:loss} guides the edges selection, the set of edges to add could be vast.
The user and item nodes of this set are described by several properties, which could support or obstruct our method.
Thus, we applied several sampling policies to narrow the set of edges (connected to user nodes in $U_D$) to be added:
%
%
%
%
%
\begin{itemize}
    \item \textbf{BM (Base)}: the base algorithm with no sampling applied.
    \item \textbf{ZN (Zero NDCG)}: selects the users with no relevant items in their top-$k$ recommendation lists, i.e. $NDCG@k = 0$.
    \item \textbf{LD (Low Degree)}: selects the $\Psi_U$\% of user nodes with the lowest degree, i.e. fewest interactions in the training set.
    \item \textbf{SP (Sparse)}: denoting a user $u$'s \emph{density} as the average popularity of the items $u$ interacted within the training set, it selects the $\Psi_U$\% of users with the lowest \emph{density} (highest \emph{sparsity}), i.e. mostly interacting with niche items.
    \item \textbf{FR (Furthest)}: selects the $\Psi_U$\% of furthest user nodes from $U_A$, where the distance from $u_D \in U_D$ is computed as the shortest paths lengths average between $u_D$ and all $u_A \in U_A$.
    \item \textbf{IP (Item Preference)}: following \cite{ijimai/DeepFair}, we estimate the extent to which an item is preferred by $U_D$; thus, $I$ is reduced by selecting the $\Psi_I$\% most preferred items by the same group.
\end{itemize}
$\Psi_U$\% and $\Psi_I$\% denote parameters to sample the user set $U$ and the item set $I$ respectively.
We fix $\Psi_U$\% = 35\% and $\Psi_I$\% = 20\%.

These policies were selected factoring in the way each demographic group interacts with the items (IP, SP), common phenomena described in recommendation literature (ZN, LD), the aggregation operation in GNNs models (LD, FR).
We distinguish between policies of type \emph{U} (ZN, LD, SP, FR) or \emph{I} (IP) if the sampling is applied on the user or item set respectively.
We also contemplated inter-group combinations between policies \emph{U} and \emph{I}, but intra-group ones are excluded not to lead to excessive reduction of the user or item set.

\section{Evaluation} \label{sec:eval}

Our experiments aim at answering the following questions:


\begin{itemize}
    \item \textbf{RQ1}: Do the edges selected by our method positively impact the recommendation fairness on the \emph{perturbation} set?
    \item \textbf{RQ2}: To what extent the unfairness is mitigated by our method in comparison with state-of-the-art procedures?
\end{itemize}


\subsection{Experiment Settings} \label{subsec:exp_settings}




\subsubsection{Data Preparation} \label{subsubsec:data_prep}

We relied on the artifacts of~\cite{10.1007/978-3-030-99736-6_37}, which performed a fairness assessment on two corpora: MovieLens 1M (ML-1M) \cite{DBLP:journals/tiis/HarperK16}, and Last.FM 1K (LFM-1K) \cite{DBLP:books/daglib/0025137}.
The advantaged groups and their representation w.r.t. to the related sensitive attribute are Males (M) (71.7\%) and Younger (Y) (56.6\%) on ML-1M, Females (F) (42.2\%) and Older (O) (42.2\%) on LFM-1K.
We extended LFM-1K with the time information to each user-artist pair as the timestamp of a given user's last interaction with a given artist's song.
Following~\cite{10.1007/978-3-030-99736-6_37}, for each data set we arranged the interactions list of each user in ascending order of recency, and split the sorted lists with a ratio 7:1:2 to include each subset in the train, validation and test set respectively.
The validation set was used (i) to select the training epoch where the model reached the highest NDCG on the non-augmented data, (ii) as the \emph{perturbation} set for our method.
During the evaluation step, the edges selected by our method were added to the training set, and, if present, removed from the other two sets.

\subsubsection{Models} \label{subsubsec:models}

Relying on Recbole \cite{DBLP:conf/cikm/ZhaoMHLCPLLWTMF21}, we adopted the following GNNs-based models to solve the top-$k$ recommendation task.

Based on an encoder-decoder architecture, \textbf{GCMC} \cite{DBLP:journals/corr/BergKW17} reconstructs the user-item relevance matrix by predicting the relevance of the missing entries in the adjacency matrix.

Leveraging high-order connectivities in the user-item interaction graph, \textbf{NGCF} \cite{DBLP:conf/sigir/Wang0WFC19} propagates embeddings in the latter by injecting the collaborative signal into the embedding process.

\textbf{LigthGCN} \cite{DBLP:conf/sigir/0001DWLZ020} is lightened to include only the neighborhood aggregation and propagate a single embedding in the graph as the weighted sum of the user and item embeddings.

We optimized the hyper-parameters under a grid search strategy.



\subsection{RQ1: Edges Augmentation Analysis} \label{subsec:rq1_edges_analysis}

\begin{table}[!t]
    \centering
    \caption{\small Mitigation performance of our method's policies: the relative difference in $\Delta$NDCG between the scores measured on the \emph{perturbation} set before and after applying each policy is reported.
    {\color{black} Negative values denote unfairness was mitigated by the respective policy, i.e. $|\Delta\text{NDCG}|$ was reduced ($-100\%$ denotes optimal mitigation)}.
    \vspace{-5mm}
    }
    \label{tab:policy_comparison}
    \resizebox{0.93\linewidth}{!}{\begin{tabular}{lll|rrrrrr}
    \toprule
        & Policy & \multirow{2}{*}{Policy} & \multicolumn{2}{c}{GCMC} & \multicolumn{2}{c}{LightGCN} & \multicolumn{2}{c}{NGCF} \\
        & Type & & \multicolumn{1}{c}{Gender} & \multicolumn{1}{c}{Age} & \multicolumn{1}{c}{Gender} & \multicolumn{1}{c}{Age} & \multicolumn{1}{c}{Gender} & \multicolumn{1}{c}{Age} \\
    \midrule
    \multirow{10}{*}{\rotatebox[origin=c]{90}{\Large \bftab ML-1M}} & & BM &        14.5\% &        15.3\% &  \hlgray{-100.0\%} &  \hlgray{-74.1\%} &          6.9\% &  10.0\% \\
    \cline{2-9}
    & \multirow{4}{*}{U} & ZN &         0.0\% &  \hlgray{-98.3\%} &   \hlgray{-99.3\%} &  \hlgray{-33.3\%} &          0.8\% &   5.5\% \\
    & & LD     &        14.5\% &        15.3\% &  \hlgray{-100.0\%} &  \hlgray{-74.1\%} &          6.9\% &  10.0\% \\
    & & SP      &         1.7\% &        13.6\% &   \hlgray{-92.4\%} &  \hlgray{-80.2\%} &         11.5\% &  10.0\%\\
    & & FR      &         5.1\% &        15.3\% &   \hlgray{-95.2\%} &  \hlgray{-81.5\%} &          5.4\% &   9.1\%  \\
    \cline{2-9}
    & I & IP & 9.4\% &       544.1\% &   \hlgray{-93.1\%} &  \hlgray{-92.6\%} &         18.5\% &   9.1\% \\
    \cline{2-9}
    & \multirow{4}{*}{U+I} & ZN+IP &  \hlgray{-53.8\%} &  \hlgray{-11.9\%} &   \hlgray{-88.3\%} &  \hlgray{-81.5\%} &  \hlgray{-100.0\%} &   8.2\% \\
    & & LD+IP  &         9.4\% &  \hlgray{-18.6\%} &   \hlgray{-93.1\%} &  \hlgray{-92.6\%} &         18.5\% &   9.1\% \\
    & & SP+IP   &         7.7\% &        10.2\% &   \hlgray{-97.2\%} &  \hlgray{-80.2\%} &          7.7\% &  15.5\%  \\
    & & FR+IP   &         9.4\% &  \hlgray{-50.8\%} &   \hlgray{-97.9\%} &  \hlgray{-66.7\%} &          4.6\% &  12.7\%\\
    \midrule
    \midrule
    \multirow{10}{*}{\rotatebox[origin=c]{90}{\Large \bftab LFM-1K}} & & BM  &  \hlgray{-93.7\%} &  \hlgray{-89.9\%} &       164.3\% &       271.0\% &  \hlgray{-0.7\%} &  \hlgray{-32.6\%} \\
    \cline{2-9}
    & \multirow{4}{*}{U} & ZN  &  \hlgray{-99.7\%} &  \hlgray{-92.4\%} &  \hlgray{-40.9\%} &  \hlgray{-97.2\%} &  \hlgray{-0.7\%} &  \hlgray{-49.9\%} \\
    & & LD     &  \hlgray{-59.2\%} &  \hlgray{-84.4\%} &       164.3\% &       271.0\% &  \hlgray{-0.7\%} &  \hlgray{-32.6\%} \\
    & & SP      &         5.8\% &  \hlgray{-93.3\%} &  \hlgray{-69.3\%} &  \hlgray{-80.9\%} &        0.2\% &  \hlgray{-32.6\%} \\
    & & FR      &  \hlgray{-95.5\%} &  \hlgray{-97.3\%} &  \hlgray{-58.6\%} &  \hlgray{-79.4\%} &  \hlgray{-0.7\%} &  \hlgray{-32.6\%} \\
    \cline{2-9}
    & I & IP  &  \hlgray{-88.4\%} &  \hlgray{-94.3\%} &   \hlgray{-6.4\%} &   \hlgray{-0.4\%} &  \hlgray{-2.4\%} &  \hlgray{-34.6\%} \\
    \cline{2-9}
    & \multirow{4}{*}{U+I}  & ZN+IP &   \hlgray{-0.3\%} &  \hlgray{-96.4\%} &   \hlgray{-3.5\%} &         0.2\% &        0.7\% &  \hlgray{-32.6\%} \\
    & & LD+IP  &     \hlgray{-88.4\%} &  \hlgray{-94.3\%} &   \hlgray{-6.4\%} &   \hlgray{-0.4\%} &  \hlgray{-2.4\%} &  \hlgray{-32.4\%} \\
    & & SP+IP   &      \hlgray{-8.9\%} &   \hlgray{-0.2\%} &   \hlgray{-3.8\%} &         0.9\% &        0.7\% &  \hlgray{-34.6\%} \\
    & & FR+IP   &         1.1\% &   \hlgray{-0.4\%} &   \hlgray{-3.8\%} &         1.3\% &        0.7\% &  \hlgray{-34.8\%} \\
    \bottomrule
    \end{tabular}}
    \vspace{-3mm}
\end{table}

\begin{figure*}
    \vspace{-3mm}
    \centering
    \resizebox{0.95\linewidth}{!}{
        \begin{tabular}{cc:c:c:c}
             & \multicolumn{4}{c}{\includegraphics[width=0.8\linewidth]{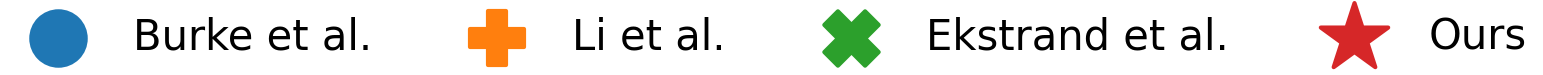}} \\
             \multirow{1}{*}[45mm]{\rotatebox[origin=c]{90}{\huge \textbf{Rel. Diff. $\Delta$NDCG}}} & \includegraphics{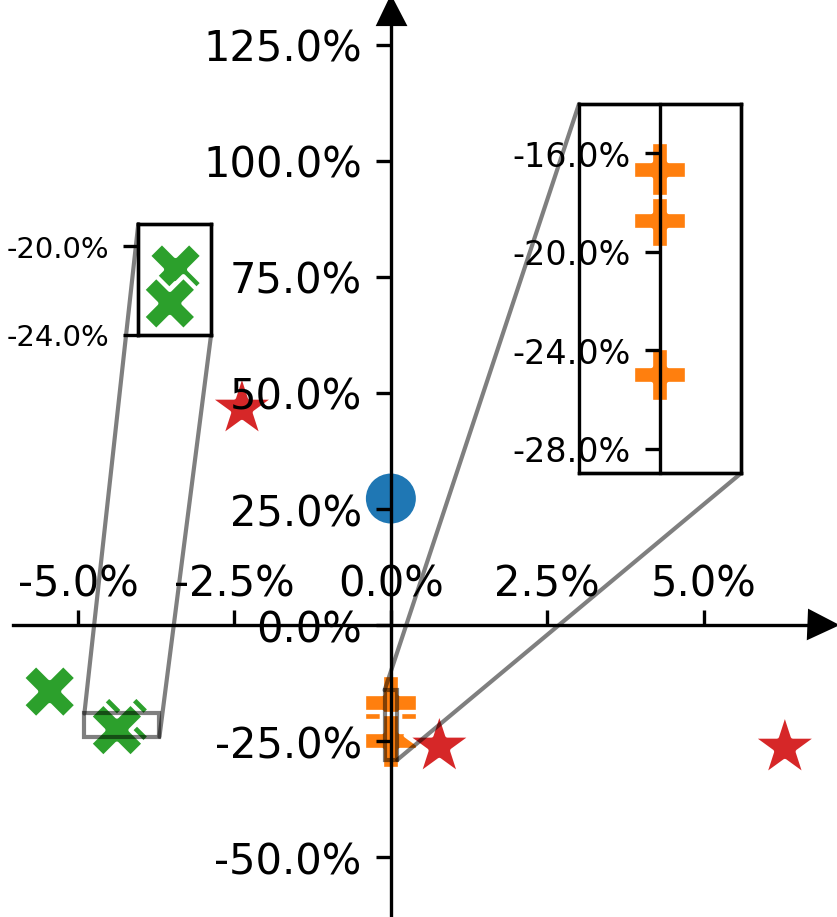} & \includegraphics{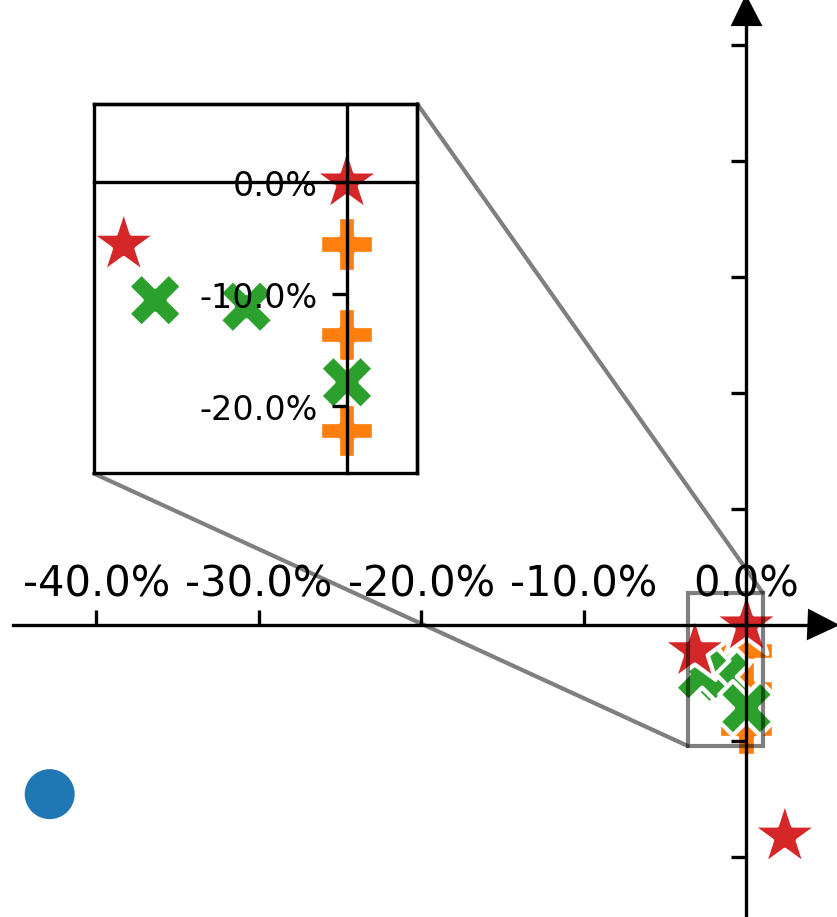} & \includegraphics{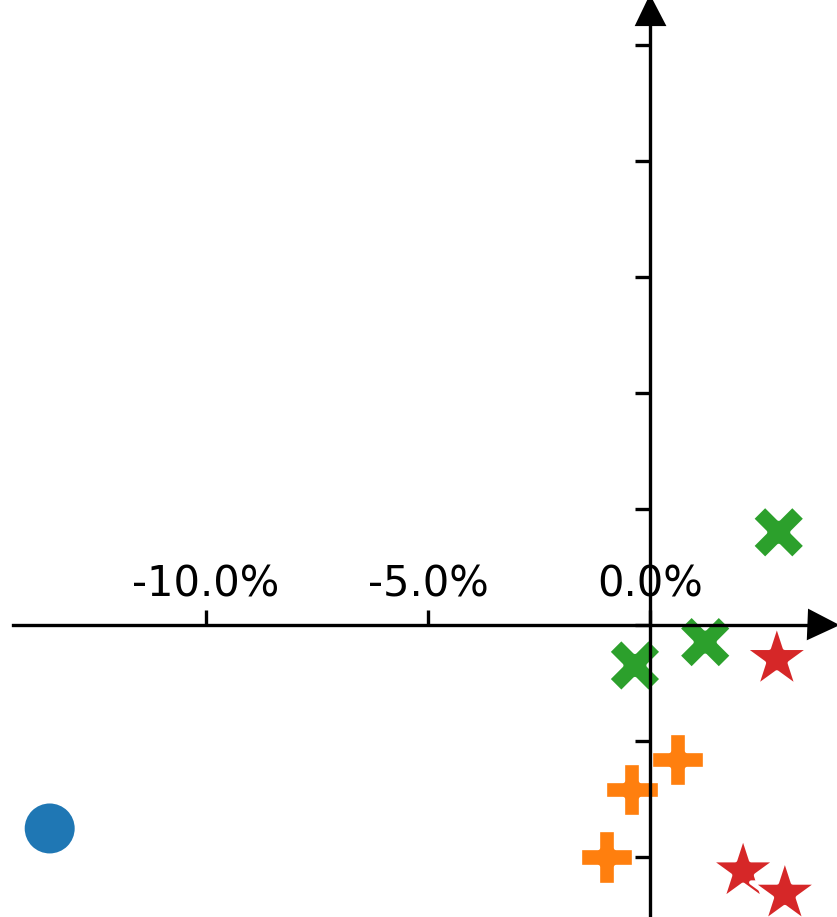} & \includegraphics{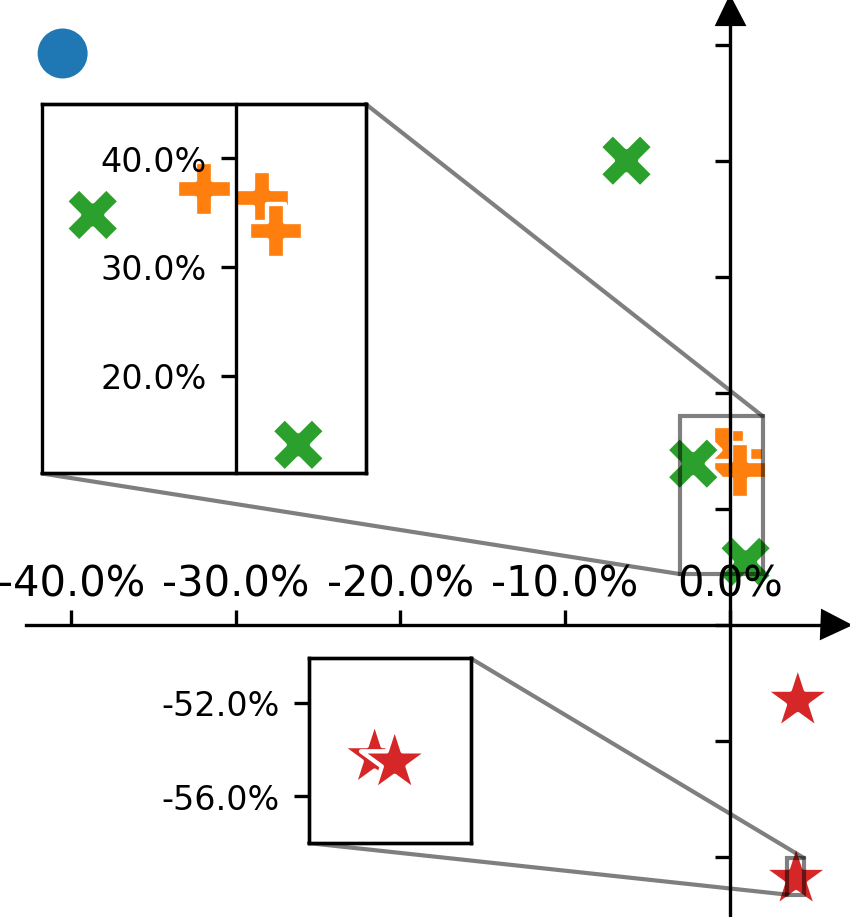} \\
        \end{tabular}
    }
        
    \bigskip

    \begin{tabular}{p{1cm}p{4cm}p{4cm}p{4cm}p{4cm}}
             & \small \textbf{Rel. Diff. NDCG} & \small \textbf{Rel. Diff. NDCG} & \small \textbf{Rel. Diff. NDCG} & \small \textbf{Rel. Diff. NDCG} \\
             &  \footnotesize (a) ML-1M Gender & $\:$ \footnotesize (b) ML-1M Age & (c) \footnotesize LFM-1K Gender & $\;$ (d) \footnotesize LFM-1K Age \\
    \end{tabular}
    \vspace{-4mm}
    \caption{\small On the x-axis (y-axis) the relative difference, denoted as \texttt{Rel. Diff.}, in recommendation utility NDCG (utility disparity $\Delta$NDCG) between the scores measured before and after each method was applied on \emph{gender} and \emph{age} groups. Multiple points per method indicate the use of multiple models for each method. Positive (negative) values on the x-axis (y-axis) denote an increment (decrement) in NDCG ($\Delta$NDCG).}
    \label{fig:relative_diff}
\end{figure*}

{\color{black} If our method successfully mitigates the model unfairness by adding the selected edges, the characteristics of the nodes composing such edges could describe a possible cause of the original unfairness (before the graph was augmented)}.
Under a given policy, the features of the sampled nodes (Section \ref{subsec:sampl_policies}) characterize the added edges.
{\color{black} Thus, Table~\ref{tab:policy_comparison} depicts the unfairness mitigation performance of all the policies to highlight which nodes characteristics affect the unfairness the most.}
Such performance is the relative difference in $|\Delta\text{NDCG}|$ between the scores measured on the \emph{perturbation} set before and after a policy was applied, where $\Delta\text{NDCG} = \overline{\text{NDCG}}_{U_D} - \overline{\text{NDCG}}_{U_A}$ is the difference between the average NDCG for $U_D$ and $U_A$.

{\color{black} Some settings are positively affected by all policies, while others can successfully be augmented only by specific policies.
Indeed, this aspect is highlighted by \texttt{ZN+IP}, the only policy mitigating unfairness across gender groups on ML-1M for GCMC and NGCF, with a noteworthy change of $-100\%$ for the latter.}
While the individual policies \texttt{ZN} (disadvantaged users with no relevant items out of the 10 recommended, i.e. $NDCG@10 = 0$) and \texttt{IP} (items mostly preferred by the disadvantaged users) could not report a similar result, their combination added interactions to the females, i.e. $U_D$, that were able to reduce the gap in NDCG between gender groups.

Some policies systematically excel more than others under the same settings, such as \texttt{ZN} across age groups on ML-1M (GCMC), and on LFM-1K (LightGCN, NGCF).
Hence, adding interactions~to the users selected by \texttt{ZN} could improve their recommendation utility.
Indeed, with an in-depth inspection, we observed that the policies reducing $\Delta$NDCG have a negligible effect on the NDCG of $U_A$, proving our approach focuses only on improving the utility of $U_D$.

Some policies consistently reducing $\Delta$NDCG regardless of the model (\texttt{ZN+IP} on ML-1M for gender ; \texttt{ZN}, \texttt{SP}, \texttt{FR} on LFM-1K for age) suggest unfairness originates at the data set level, but other policies not working for different models (\texttt{ZN}, \texttt{FR+IP} on ML-1M for age ; \texttt{LD} on LFM-1K for age) underline that the bias is model-dependent.

\subsection{RQ2: Mitigation Procedures Comparison} \label{subsec:rq2_mitig_comparison}

\begin{table}[!t]
    \centering
    \caption{\small Recommendation utility (NDCG) and utility disparity ($\Delta$NDCG) after applying each method on user groups. For each setting, the best and second best scores are in bold and italic respectively.}
    \label{tab:mitig_comparison}
    \vspace{-3mm}
    \resizebox{\linewidth}{!}{
    \begin{tabular}{ll|rr|rr|rr|rr}
    \toprule
              &     & \multicolumn{4}{c|}{\Large \bftab ML-1M} & \multicolumn{4}{c}{\Large \bftab LFM-1K} \\
              &     & \multicolumn{2}{c|}{NDCG $\uparrow$} & \multicolumn{2}{c|}{$\Delta$NDCG $\downarrow$} & \multicolumn{2}{c|}{NDCG $\uparrow$} & \multicolumn{2}{c}{$\Delta$NDCG $\downarrow$} \\
              Paper & Model & \multicolumn{1}{c}{Gender} & \multicolumn{1}{c|}{Age} & \multicolumn{1}{c}{Gender} & \multicolumn{1}{c|}{Age} & \multicolumn{1}{c}{Gender} & \multicolumn{1}{c|}{Age} & \multicolumn{1}{c}{Gender} & \multicolumn{1}{c}{Age} \\
    \midrule
    \cite{DBLP:conf/fat/BurkeSO18} & SLIM-U &     0.084 &  0.048 &  {\scriptsize \^{}}0.028 & \ittab {\scriptsize \^{}}0.014 &  0.301 & 0.207 &  {\scriptsize \^{}}0.072 &  {\scriptsize \^{}}-0.145  \\
    \midrule
    \cite{DBLP:conf/www/LiCFGZ21} & BiasedMF &  0.112 &  0.112 & \bftab {\scriptsize \^{}}0.013 &  {\scriptsize \^{}}0.017 &  0.245 & 0.247  &  {\scriptsize *}-0.049 &     {\scriptsize *}-0.060 \\
              &   NCF &     0.120 & 0.120 &  {\scriptsize \^{}}0.015 &   {\scriptsize \^{}}0.019 &  0.202  &   0.203  & \bftab  -0.023 &     -0.048 \\
              &   PMF & 0.123 & 0.123 & {\scriptsize \^{}}0.015 & {\scriptsize \^{}}0.021 & 0.164 &  0.164 & {\scriptsize *}-0.049 & {\scriptsize \^{}}-0.044 \\
    \midrule
    \cite{DBLP:conf/fat/EkstrandTAEAMP18} &  ItemKNN & \ittab 0.134 & \bftab  0.138 &  {\scriptsize \^{}}0.030 &  {\scriptsize \^{}}0.024 &  0.286 &  0.269  & {\scriptsize *}-0.116 &        \bftab  0.020 \\
              &   TopPopular &  0.104 & 0.107 &  {\scriptsize \^{}}0.030 &  {\scriptsize \^{}}0.034 &   0.321 &  0.315 & {\scriptsize *}-0.102 &   -0.050 \\
              &   UserKNN &   0.131 &  \ittab   0.137  & {\scriptsize \^{}}0.024 &  {\scriptsize \^{}}0.023 &  \bftab  0.411 & 0.397 &  {\scriptsize \^{}}-0.106  &    -0.031 \\
    \midrule
    Ours &  GCMC   &  0.123 &   0.122 &  {\scriptsize \^{}}0.022 &  {\scriptsize \^{}}0.017 &  0.392 & 0.399 & \ittab -0.024  &  -0.039 \\
                 &   LightGCN & \bftab 0.135 & 0.130 & \ittab -0.014 & \bftab {\scriptsize \^{}}0.012 &  \ittab 0.409 &  \bftab 0.413 &   -0.030 & \ittab -0.030 \\
                 &   NGCF &  0.130 & 0.129 &  {\scriptsize \^{}}0.017 &  {\scriptsize \^{}}0.023 & 0.398 & \ittab 0.403 &  {\scriptsize *}-0.077  &  -0.062 \\
    \bottomrule
    \end{tabular}
    }
    \vspace{-3mm}
\end{table}

In this section, we evaluate the trade-off between the recommendation utility and the unfairness mitigation performance of our method in comparison with SOTA algorithms.
Based on the similarity to our evaluation protocol, we relied on the framework shared by~\cite{10.1007/978-3-030-99736-6_37}\footnote{Experiments on LFM-1K were re-run to match our splitting strategy.} and compared the mitigation procedures used for top-$k$ recommendation with our method.
Given our focus on the mitigation task, we only considered models reporting high utility levels\footnote{We discarded LBM, STAMP, FunkSVD since they reported a NDCG lower than half of the best models one (ItemKNN for ML-1M, UserKNN for LFM-1K) for both data sets.}, which could effectively solve the recommendation task and reflect existing biases as in real-world scenarios.
The GNN-based recommendation systems used with our algorithm satisfy this property.
The following results regarding our method pertain to the policies that reported the lowest $|\Delta\text{NDCG}|$ on the \emph{perturbation} set (Table~\ref{tab:policy_comparison}).

Figure~\ref{fig:relative_diff} highlights the extent to which each method affected the recommendation utility in NDCG (x-axis) and the disparity in the latter between user groups (y-axis).
Our approach reports the best mitigation performance, given by the points labeled as \texttt{Ours} being systematically the lowest ones on the y-axis.
{\color{black} Moreover, except for GCMC on ML-1M, our algorithm systematically reported a positive (right side of x-axis) or negligible impact on the recommendation utility, whereas the other methods decreased it in several settings.}
In Table~\ref{tab:mitig_comparison} we report the resulting levels of utility (NDCG) and fairness ($\Delta$NDCG) after each algorithm was applied.
Determining which setting is the best one depends on a specific application requirements and to what extent fairness is relevant.
{\color{black} In terms of recommendation utility, our approach made the GNN-based models among the most effective in various settings, and it led to utility disparity levels significantly lower than the other systems reporting a high NDCG, e.g. LFM-1K on gender groups.}
Hence, our algorithm demonstrates greater reliability in mitigating unfairness and in improving recommendation utility than the other methods.

\section{Conclusions}


In this paper, we proposed an augmentation method that leverages explanation techniques to mitigate consumer unfairness in recommendations generated by a GNN-based system.
Our experiments show our technique as more reliable to mitigate unfairness than SOTA algorithms, and as able to increase the overall recommendation utility.
Analyzing the augmented graph, we discovered that confining the algorithm to the disadvantaged user nodes who received no relevant items in their recommendation list positively affects the utility of the latter.
However, {\color{black} the augmentation had a limited impact on deep GNNs (GCMC, NGCF), primarily due to the diminished influence of the graph in the prediction process.
It is also unclear whether our approach could improve the disparity in other metrics, given that it solely focuses on the NDCG.}
Future works will consider new policies, objective functions, and the adoption of our method to other models, not necessarily based on GNNs.

\bibliographystyle{ACM-Reference-Format}
\bibliography{sample-base}

\end{document}